\documentclass[prd,aps]{revtex4}

\usepackage{graphicx}
\usepackage{epsfig}

\begin{document}

\markboth{C. Hays, A. Kotwal, and O. Stelzer-Chilton}
{New Techniques in the Search for Z' Bosons and Other Neutral Resonances}

\title{New Techniques in the Search for Z' Bosons and Other Neutral Resonances\\
}

\author{\footnotesize Christopher Hays}

\affiliation{University of Oxford, \\
Oxford, OX1 3RH,
United Kingdom\\
hays@physics.ox.ac.uk}

\author{Ashutosh Kotwal}

\affiliation{Duke University, \\
Durham, NC 27708, USA\\
kotwal@phy.duke.edu}

\author{Oliver Stelzer-Chilton}

\affiliation{TRIUMF, \\
Vancouver, British Columbia V6T 2A3, Canada\\
stelzer-chilton@triumf.ca}

\begin{abstract}
The search for neutral resonances at the energy frontier has a long and 
illustrious history, resulting in multiple discoveries.  The canonical search 
scans the reconstructed invariant mass distribution of identified fermion 
pairs.  Two recent analyses from the CDF experiment at the Fermilab Tevatron 
have applied novel methods to resonance searches.  One analysis uses simulated 
templates to fit the inverse mass distribution of muon pairs, a quantity with 
approximately constant resolution for momenta measured with a tracking detector.
The other analysis measures the angular distribution of electron pairs 
as a function of dielectron mass, gaining sensitivity over a probe of the mass 
spectrum alone.  After reviewing several models that predict new neutral 
resonances, we discuss these CDF analyses and potential future applications.

\keywords{Z' boson; graviton; sneutrino; CDF; Tevatron.}
\end{abstract}

\maketitle

\section{Introduction}	

Searches for neutral resonances have historically brought major breakthroughs 
by either confirming important predictions or discovering unexpected particles.  
The 1974 discovery of the $J/\psi$ meson \cite{jpsi} as a $c\bar{c}$ bound state 
confirmed the GIM mechanism \cite{gim} for preventing flavor-changing neutral 
currents, and the discovery of the $Z$ boson \cite{z0} confirmed the gauge 
unification of the electromagnetic and weak forces \cite{SM}.  Meanwhile, the 
discovery of the upsilon \cite{upsilon} was completely unexpected, and increased 
the number of known fermion generations to three.

Turning to the future, there are reasons to expect the next important particle 
physics discovery will be a neutral resonance.  In addition to the well-motivated 
Higgs boson \cite{higgs} of the standard model (SM), there are many new resonances 
predicted by proposed extensions to the standard model.  These extended theories can 
address unexplained features of the SM, such as: the lack of gauge unification and the 
hierarchy between the electroweak and Planck scales (through supersymmetry \cite{susy} 
or the presence of extra dimensions \cite{led}); and parity violation and light 
neutrino masses (through an additional SU(2)$_{\rm R}$ gauge symmetry \cite{lrsym}, 
which has weak couplings to right-handed fermions).  

The most sensitive direct searches for neutral resonances at high mass come from 
Tevatron $p\bar{p}$ collision data.  Future searches in $pp$ collisions from the 
Large Hadron Collider (LHC) \cite{lhc} will increase the probed mass range.  As 
larger datasets with higher energies are studied, enhancements to the search 
strategy can improve sensitivity and facilitate the analysis.  Such enhancements 
have been developed and applied to searches for new resonances in the CDF 
dimuon \cite{CDFDimuon} and dielectron \cite{cdfafb} data.

\section{Models Containing Neutral Resonances}

A neutral resonance decaying to fermion pairs can have intrinsic spin equal 
to 0, 1, or 2.  No fundamental scalar particle has yet been observed, though 
the SM requires one in the form of a Higgs boson.  Beyond the SM, there 
could be multiple Higgs bosons with varying properties \cite{nonsmhiggs}.  In 
supersymmetric models, there are spin-0 partners to fermions that could be 
produced as resonances in $p\bar{p}$ or $pp$ collisions \cite{snuprod}.  Any 
model with an additional U(1) gauge group will have a new spin-1 gauge boson, 
generically referred to as a $Z'$ boson \cite{firstz',z'review}.  Models of 
extra dimensions at the electroweak scale predict spin-2 graviton 
resonances \cite{rs,gravitons}.

\subsection{Sneutrino Production in Hadron Collisions}

To remove the fine-tuning of the Higgs boson mass, the scale of supersymmetry should 
be of the same order as the electroweak scale, making the discovery of supersymmetry 
likely at the Tevatron or LHC (should it exist).  The supersymmetric partner to 
the neutrino has no electromagnetic charge and is thus a candidate for production 
as a neutral resonance.    

Resonant sneutrino production would violate $R$-parity, a multiplicative quantum 
number that is +1 for matter and -1 for supersymmetric matter.  The violation of 
$R$-parity implies that the lightest sparticle is not stable, potentially removing 
it as a candidate for dark matter.  However, if the lightest sparticle has 
sufficiently small couplings to give it a lifetime on the order of the age of the 
universe, it can still be a dark-matter candidate \cite{dm}.

\begin{figure}[!htbp]
\begin{center}
\epsfysize = 1.5in
\epsffile{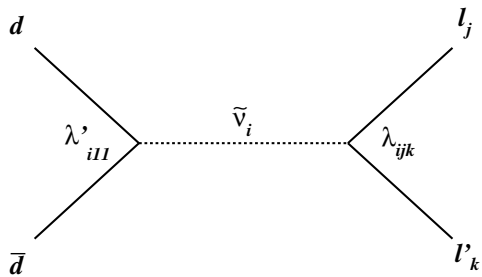}
\caption{The Feynman diagram for resonant sneutrino production at a hadron collider. }
\label{snuprod}
\end{center}
\end{figure}

Proton decay limits require at least one set of $R$-parity-violating terms to be 
vanishingly small \cite{pdecay}.  This can be accomplished by imposing a ``baryon 
parity" that conserves baryon number and suppresses proton decay more than 
$R$-parity conservation \cite{bparity}.  With baryon parity there are two sets of 
$R$-parity-violating Yukawa terms in the Lagrangian, both of which are consistent 
with proton decay limits and relevant for sneutrino production and decay at a 
hadron collider:
\begin{equation}
{\cal L}_{R \! \! \! /} = \lambda_{ijk} L_i L_j e^c_k + \lambda'_{ijk} L_i Q_j d^c_k,
\end{equation}

\noindent
where $Q$ ($d$) and $L$ ($e$) are SU(2)$_{\rm L}$ doublet (singlet) superfields.  
As shown in Fig.~\ref{snuprod}, the first term governs sneutrino decay to leptons 
and the second term governs sneutrino production in hadron collisions.  Because of 
the $d^c_k$ superfield in the production term, only-down type quark interactions 
produce sneutrinos, with $\lambda'_{i11}$ the most relevant coupling at the Tevatron 
and LHC (due to parton distributions in the proton).

The total width of a new resonance is an important parameter in a search, in 
particular relative to the detector mass resolution.  The partial width of a 
given sneutrino decay is \cite{snuwidth,units}:
\begin{equation}
\Gamma(\tilde{\nu}_i \rightarrow f_j f_k) = \frac{c_{jk}}{16\pi} \lambda^2 m_{\tilde{\nu}_i} ,
\end{equation}

\noindent
where $c_{jk}$ is a color factor, and $\lambda$ ($=\lambda_{ijk}$ or 
$\lambda'_{ijk}$) is the coupling to the final-state $f_j f_k$.  The width is 
fairly narrow; for example, if only $\lambda'_{i11}$ and one $\lambda_{ijk}$ 
are large, then for respective values of 1/2 and 1, the width is 3.5\% of 
the mass (this can be compared to the $Z$ boson, whose width is 2.8\% of 
its mass \cite{z0width}).

A range of indirect limits exist on $\lambda'_{i11}$ \cite{snuindirect}, and 
generally depend on other supersymmetric parameters.  A typical set of limits 
comes from the ratio of $\Gamma(\pi \rightarrow e \nu)/\Gamma(\pi \rightarrow \mu \nu)$ 
or $\Gamma(\tau \rightarrow \pi \nu)/\Gamma(\pi \rightarrow \mu \nu)$:  
$\lambda'_{i11} < a_i m_{\tilde{d}}/$TeV, where $a_1 = 0.26$, $a_2 = 0.59$, and 
$a_3 = 1.2$.    For reasonable values of $m_{\tilde{d}}$ ($0.2 - 1$ TeV), these 
limits allow significant production rates with a relatively narrow decay width.

\subsection{$Z'$ Vector Bosons}

Many models predict a new electroweak-scale U(1) gauge symmetry 
\cite{techni,lh,nmssm,e6review}, which would have an associated $Z'$ gauge 
boson.  A useful test model is a superstring-inspired grand unified theory with 
E$_8 \times$E$'_8$ gauge structure \cite{e8e8}.  In this model, the E$'_8$ group 
is a hidden sector that breaks supersymmetry, and the E$_8$ group is broken to 
E$_6 \times$SU(3) by the compactification of extra dimensions \cite{e6}.  Each 
generation of matter particles fits in a fundamental 27 representation of E$_6$; 
thus, before E$_6$ is broken, each SM generation is just a single field 
distinguished by its E$_6$ charge.  

The range of options for breaking E$_6$ to the SM gauge structure allows a variety 
of phenomena \cite{z'review}.  A symmetry-breaking proceeding through SO(10)$\times$U(1) 
to SU(4)$_{\rm C} \times$SU(2)$_{\rm L} \times$SU(2)$_{\rm R} \times$U(1)$_{\rm B-L}$ 
restores parity conservation and provides for the seesaw mechanism for small neutrino 
masses.  Alternatively, the breaking can proceed through SO(10)$\times$U(1)$_{\psi}$ 
to SU(5)$\times$U(1)$_{\chi} \times$U(1)$_{\psi}$, producing two new U(1) gauge 
groups.  At a high mass scale, the SU(5) can be broken to the SM gauge 
groups and one of the extra U(1) gauge groups can be broken, potentially leaving 
one non-SM ${\rm U(1)'}$ at the electroweak scale.  Taking this ${\rm U(1)'}$ to 
be a linear combination of U(1)$_{\psi}$ and U(1)$_{\chi}$,
\begin{equation}
{\rm U(1)'} = {\rm U(1)}_{\psi} \cos\theta + {\rm U(1)}_{\chi}\sin\theta,
\end{equation}

\noindent
a generic ${\rm U(1)'}$ can be expressed in terms of $\theta$ \cite{u1mixing}.

Scanning the $\theta$ parameter space gives models with distinct phenomena.  The 
secluded ${\rm U(1)'}$ ($\theta = \pi - \tan^{-1}\sqrt{27/5}$) is mediated by a 
$Z'_{sec}$ boson whose mass results from the vacuum expectation value of a scalar 
field with no SM charge \cite{secluded}.  For $\theta = -\tan^{-1}\sqrt{1/15}$, 
the right-handed neutrino has no charge in the extended gauge group 
(SU(3)$_{\rm c} \times$SU(2)$_{\rm L} \times$U(1)$_{\rm Y} \times$U(1)$_{\rm N}$), 
and is thus sterile \cite{u1N}.  The breaking of E$_6$ directly to the SM groups 
plus U(1)$_{\eta}$ corresponds to $\theta = \tan^{-1}\sqrt{3/5}$.  If 
the breaking proceeds through an extra SU(2)$_{\rm I}$ group (instead of 
SU(2)$_{\rm R}$), then the $W$' and $Z'$ bosons of the new group have zero 
electromagnetic charge.  

More general classes of models have also been considered, with the constraint 
of anomaly cancellation to produce a consistent theory \cite{nnlo}.  Under the 
assumption of the SM Higgs mechanism for generating fermion masses, a general 
class of models has ${\rm U(1)'}$ charge $B - xL$, where $B$ ($L$) is baryon 
(lepton) number (and the right-handed neutrino charge is fixed to -1).  Allowing 
for non-SM mass generation but considering only SM particles for anomaly 
cancellation gives ${\rm U(1)'}$ charges 1/3, $x/3$, $(2-x)/3$, -1, $-(2+x)/3$, 
and $(-4+x)/3$ for the states $q_L$, $u_R$, $d_R$, $l_L$, $e_R$, and $\nu_R$, 
respectively.  This model is referred to as $q+xu$ and includes the case of $B-L$ 
symmetry for $x = 1$.  Two additional model classes arise when two non-SM fermions 
are added to the theory.  One is referred to as $d - xu$ and has charges of 0, 1/3, 
and $-x/3$ for $q_L$, $d_R$, and $u_R$, respectively.  The other, $10+x\bar{5}$,
 has fermions in the 10 representation of the SU(5) grand unified group with the 
$U(1)'$ charge 1/3, and fermions in the $\bar{5}$ representation with charge $x/3$.

In general, couplings of the new $Z'$ boson to SM particles are smaller than those 
of the $Z$ boson in the SM.  However, the new $Z'$ boson could decay into the non-SM 
particles that are part of the 27 representation of E$_6$.  If decay to all of these 
particles is possible, the $Z'$ boson width could be $\approx$5\% of its 
mass \cite{z'width}.  Even in this extreme case, the $Z'$ boson would appear as a 
narrow resonance.

\subsection{Graviton Resonances}

It has been suggested that the apparent difference between the scales of gravity 
and electroweak symmetry-breaking is due to the presence of at least one unobserved 
spatial dimension \cite{led,rs}.  The spread of the gravitational field into the 
extra dimension(s) weakens the strength of gravity in the observed dimensions.  
Closing the gap between the electroweak and Planck scales requires either the number 
or the size of the extra dimensions to be large, if they are flat.  

Recently, Randall and Sundrum have proposed a model that removes the scale hierarchy
using one small extra dimension \cite{rs}.  This can be accomplished with a warped 
dimension separating the SM brane from the gravity brane, resulting in a metric 
of the form:
\begin{equation}
ds^2 = e^{-2kr\phi} \eta_{\mu\nu} dx^{\mu} dx^{\nu} - r^2 d\phi^2,
\end{equation}

\noindent
where $r$ is the compactification radius, and $k^2$ and $\phi = [-\pi, \pi]$ are the 
spacetime curvature and coordinate in the extra dimension, respectively.  The curvature 
is of the order of $M_{Pl}^2$, where $M_{Pl} = G_N^{-1/2} \sim 10^{19}$ GeV is the 
Planck scale on the 4-dimensional spacetime and $G_N$ is the Newtonian gravitational 
constant.  For a string theory with ${\cal O}$(1) couplings, 
$k/M_{Pl} \sim 0.01$ \cite{gravitons}. 

Due to the spacetime warping, distances are exponentially larger on the gravity brane, 
resulting in a large gravitational field flux on this brane.  The gravitational force 
for an observer on the SM brane appears as:
\begin{equation}
F \sim m_1 m_2 / (M_{EW}^2 e^{2kr\pi} R^2), 
\end{equation}

\noindent
where $R$ is the distance between masses $m_1$ and $m_2$ in the three large spatial 
dimensions.  Thus, $kr \sim 12$ reproduces the observed weakness of gravity and there 
are no large hierarchies in the model.  In terms of the gravitational quantum, the 
wave function of the massless graviton state is localized on the gravity brane and 
exponentially suppressed on the SM brane.

Graviton excitations are localized on the SM brane and are thus expected to have mases 
$m_n$ at the electroweak scale,
\begin{equation}
m_n = k x_n e^{-kr\pi},
\end{equation}

\noindent
where $x_n$ are ${\cal O}(1)$ roots of a Bessel function and $k e^{-kr\pi}$ is of the 
order of the electroweak scale.  The resonance width is proportional to $(k/M_{Pl})^2$, 
and is less than a few percent for $k/M_{Pl} \leq 0.1$.  

\section{Collider Searches for Neutral Resonances}

The most stringent direct limits on new neutral resonances come from searches at 
the Tevatron.  Run II searches at the CDF and D0 experiments have probed resonance 
decays to pairs of electrons \cite{CDFDielectron,diele,dielemu,d0epho,d0emupho,sleuth}, 
muons \cite{dielemu,d0emupho,sleuth}, taus \cite{sleuth,d0htautau,cdfhtautau,ditau}, 
light quarks \cite{sleuth,dijet}, top quarks \cite{sleuth,ditop}, gluons \cite{sleuth,dijet}, 
photons \cite{d0epho,d0emupho,sleuth,diphoton}, and $W$ \cite{diw} and $Z$ \cite{diz} 
bosons.  The most sensitive searches use long-lived final-state particles, while other 
searches cover parameter space where couplings to long-lived particles are suppressed.

Searches for a resonance decaying to a pair of stable particles typically probe the 
reconstructed invariant mass distribution for evidence of a narrow peak, with the 
peak width determined primarily by detector resolution.  Because detector resolutions 
increase with increasing mass, the expected peak width also increases.  This 
complication usually results in search windows that change as a function of mass, with 
the window causing some loss of acceptance.  A recent CDF search \cite{CDFDimuon} uses 
templates to fit the full dimuon inverse invariant mass spectrum for new resonances, 
avoiding the acceptance loss from a search window.  In addition, the inverse mass 
distribution has approximately constant resolution, simplifying the search.

A complement to the invariant mass distribution is the angular distribution of the 
final-state particles, which can be used to separate a signal from the SM background and 
to determine the spin of the new resonance.  CDF has performed a search in the dielectron 
final state using the $\cos\theta^*$ distribution \cite{cdfafb}, where $\theta^*$ is the 
angle between the electron and the incoming quark in the boson rest frame \cite{costheta}.  

\subsection{CDF Dimuon Search}

Currently, the CDF dimuon analysis of 2.3 fb$^{-1}$ of $p\bar{p}$ collision data is the 
most sensitive search for neutral resonances over most of the probed parameter space.  The 
search uses a parametric simulation to model the detector response and resolution for muons, 
calibrated using known resonances.  After normalizing the SM inverse mass spectrum to the 
$Z$ boson peak, the data are fit as functions of the number of new-neutral-resonance events 
above background and the resonance pole mass.  No statistically significant excess above the 
background is observed, and limits are set for the various test models.

\subsubsection{Detector Alignment and Calibration}

In the CDF search, muon momenta are measured with the central outer tracker (COT) \cite{cot}, 
a wire drift chamber embedded in a 1.4 T magnetic field covering $|\eta| < 1$ and radii 43 cm 
to 133 cm \cite{conventions}.  The reconstructed tracks are constrained to originate from the 
time-averaged transverse beam collision coordinate, significantly improving momentum 
measurement resolution.  Calorimeters and a muon detector system at large radii from the 
beam line are used for muon identification and event triggering.  

To minimize bias and optimize the detector resolution (and thus the statistical significance 
of a narrow resonance), a detailed alignment of the COT is performed \cite{CDFWMass}.  The 
alignment uses cosmic-ray muons reconstructed as a single track through both sides of the 
nominal collision point \cite{cosmicnim}.  With 96 radial wire layers in the COT, any 
given two-sided track has up to 192 measurement points.  The 96 layers are divided into 8 
superlayers of 12 wires each, with each superlayer containing enough 2 cm wide cells to cover 
the azimuth.  The first stage of the alignment allows a rotation and a shift of each cell, 
such that the mean residual of hits in any given cell is statistically consistent with zero 
(with a precision of a few microns).

After the individual cell alignment, a global correction to the wire shape 
between endplates is derived as a function of $\phi$ and radius.  The shape 
is determined by the gravitational sag from the weight of the wire, and by 
the electrostatic deflection from the local electric field.  The nominal 
correction to the wire shape due to these effects is further modified with 
an empirical correction function derived from measured biases between the 
two separately fit sides of the cosmic-ray tracks.  

A final correction to the track curvature is applied after the track 
reconstruction.  The correction is derived from the difference in the 
ratio of calorimeter energy to track momentum for positrons and electrons, 
as functions of $\phi$ and $\cot\theta$ \cite{conventions}.

The momentum scale is calibrated by tuning the measured $J/\psi$, $\Upsilon$, 
and $Z$ boson masses to their precisely known values.  Individual hit 
resolutions of 150 microns are determined from the observed width of the 
$\Upsilon\rightarrow\mu\mu$ peak, consistent with hit residuals of muons 
from $Z$ boson decays.  The transverse beam profile is modelled as a 
gaussian with a size set by the observed width of the $Z\rightarrow\mu\mu$ 
mass peak measured with beam-constrained tracks.  

\subsubsection{Inverse Mass Scan}

Muon momenta transverse to the beam line are determined from a 
measurement of the reconstructed track curvature $c$.  The Lorentz force 
$q\vec{v}\times\vec{B}$ causes a helical trajectory of the muon, 
resulting in a transverse momentum:
\begin{equation}
p_T = eBR,
\end{equation}

\noindent
where $R$ is the radius of curvature and $B$ is the magnitude of the magnetic 
field.

The resolution on the track curvature ($c \equiv R^{-1}/2$ at CDF) can 
be derived from the individual hit resolutions.  Taking as an example 
a muon produced with no impact parameter at $\phi = 0$, its position 
at a given radius $r$ is $(r\cos\phi, r\sin\phi)$.  Using the equation of 
the track circle, $x^2 + (y - R)^2 = R^2$, the curvature is $c = \sin\phi/r$ 
and its resolution due to a measured hit is $\delta c = \cos\phi \delta \phi/r$.  
The partial derivative of the curvature with respect to the hit resolution 
$\delta D = r\delta \phi$ is then 
\begin{equation}
\delta c / \delta D = (1 - c^2 r^2)^{1/2} / r^2.
\end{equation}

\noindent
For muons with small curvature (large momentum), the resolution is effectively 
independent of curvature and improves with the square of the detector radius.

Defining the reconstructed muon energies to be $E_1$ and $E_2$, and their 
opening angle to be $\Theta$, the measured mass can be expressed as: 
\begin{equation}
m = [2 E_1 E_2 (1 - \cos\Theta)]^{1/2}.
\end{equation}

\noindent
The dominant contribution to the mass resolution comes from the momentum 
measurement, since the angular resolution is negligible by comparison.  
A high-mass resonance is predominantly produced with a relatively small 
transverse boost, resulting in muons with similar transverse momenta.
Then, $m \propto p_T$, or $1/m \propto c$.  Thus, a new narrow resonance 
would have an approximately constant width in the reconstructed $1/m$ 
distribution of central muons.

The search in a constant-width variable simplifies the analysis.  A 
distribution with uniform binning can be visually scanned for resonances.
Taking bin widths sufficiently narrow with respect to the resolution allows 
a template fit for a resonance centered on each bin, with the step size an 
equal fraction of the peak width throughout the distribution.  This procedure 
optimizes the scan of the $1/m$ distribution.

At CDF, the inverse mass resolution is 17\% TeV$^{-1}$, with an additional 
contribution from multiple scattering at low mass.  At 100 GeV, the 
total resolution is about 30\% smaller than the intrinsic width of the $Z$ 
boson.  Thus, the width of the new resonance could noticeably broaden the 
peak at low mass.  However, the relative resolution increases linearly with 
increasing mass, while the relative instrinsic width remains constant, so 
for most resonances the detector resolution will dominate above a few hundred 
GeV.

In the CDF analysis, the search region is 35 bins of $m^{-1} < 10$ TeV$^{-1}$, 
resulting in an expected peak width of about 3 bins due to detector resolution.  
The 70-100 GeV mass range is used to normalize the expected background, 
effectively removing systematic uncertainties due to luminosity, background 
cross section, and trigger and muon identification efficiencies.  For each 
probed inverse mass bin, a template of combined signal and background is 
compared to the data to determine the number of signal events that maximizes 
the log Poisson likelihood.  

The template fit adds acceptance outside of the usual mass window, particularly 
at masses near the kinematic threshold.  While the template neglects 
interference between the new resonance and $Z$ and $\gamma$ bosons, any 
interference has a small effect on the search because the resonance is narrow 
and has a small cross section \cite{nnlo}.

\subsubsection{Backgrounds}

Dimuon production at a hadron collider occurs predominantly through the 
Drell-Yan process of $Z/\gamma^*$ production.  This process has been 
calculated at next-to-next-to-leading order (NNLO) in $\alpha_s$ \cite{nnlo} 
and next-to-leading order (NLO) in $\alpha_{EW}$ \cite{nlo} at the Tevatron, 
and is thus well understood theoretically.  Since the CDF analysis normalizes 
the background to the $Z$ boson peak, the shape of the inverse mass 
distribution is the important theoretical input.  Over the probed mass 
region, the ratio of NNLO to leading order (LO) predictions have an 
$\approx 10\%$ variation.  The difference between NLO and NNLO predictions 
is taken as a systematic uncertainty and increases from zero at 91 GeV (the 
normalization mass) to 9\% for a mass of 1 TeV.  A more important systematic 
uncertainty at 1 TeV comes from uncertainties on the parton distribution 
functions (PDFs).  At this mass, the valence (anti-)quarks must have a large 
fraction of the (anti-)proton's momentum.  The uncertainty obtained from a 
comparison of CTEQ \cite{cteq} PDFs fit using LO and NLO inputs increases with 
mass to approximately 16\% at 1 TeV.  No NLO $\alpha_{EW}$ correction is 
applied, and a 3\% uncertainty at 1 TeV is incorporated to cover its neglect.

A background relevant at high mass arises from hadrons decaying to muons 
in the COT, where the track is misreconstructed as a nearly straight line.  
The misreconstruction occurs via a track kink at the decay vertex or 
other incorrectly assigned hits in the inner superlayers.  This background 
is reduced by requiring each muon's COT hit pattern and track fit $\chi^2$ 
to be consistent with a well-measured track resulting from the collision 
vertex.  The small residual background is estimated using dimuons with the 
same charge, which are assumed to arise from misidentification or other 
non-prompt sources.

$W$ boson pairs produced either directly or through top quark decays
contribute to the dimuon sample when both $W$ bosons decay to 
$\mu\nu$.  These backgrounds have been calculated to NLO in 
$\alpha_s$ \cite{ww,tt} and are only relevant at high mass.

A final potential background arises from cosmic-ray muons passing 
through the detector.  These muons can have high energies and thus 
contribute to the background at high mass.  At CDF the cosmic-ray 
background is effectively eliminated by the two-sided track fit used 
to identify cosmic-ray muons in the COT \cite{cosmicnim}.

\subsubsection{Signal Cross Sections and Acceptance}

Cross sections for sneutrino production have been calculated at NLO in 
$\alpha_s$ \cite{snuxsec}.  If there is CP conservation in the sneutrino 
sector, the anti-sneutrino and sneutrino cross sections will be the 
same; this assumption is made in the CDF analysis.  $Z'$-boson and 
Randall-Sundrum-graviton cross sections are determined from 
{\sc pythia} \cite{pythia} with an NNLO $\alpha_s$ correction factor 
applied \cite{nnlo}, and the appropriate couplings are used for the 
various E$_6$ models \cite{e6couplings}.

In the CDF analysis, acceptance is calculated as a function of spin 
and inverse mass.  For 1 TeV resonances, which are produced with 
relatively little longitudinal momenta, the acceptance for observing 
two central muons is $\approx 40\%$.  At 100 GeV, the larger 
average longitudinal boost reduces the acceptance for the CDF central 
muon and tracking chambers, resulting in an acceptance of $\approx 15\%$.  
A relative uncertainty of 3\% is estimated from comparisons of the 
parametric and {\sc geant}-based detector simulations.  A study of 
$Z\rightarrow\mu\mu$ events shows the momentum dependence of the muon 
identification efficiency to be well modelled by the simulation.

\subsubsection{Results and Limits}

\begin{figure}[!h]
\centerline{\psfig{file=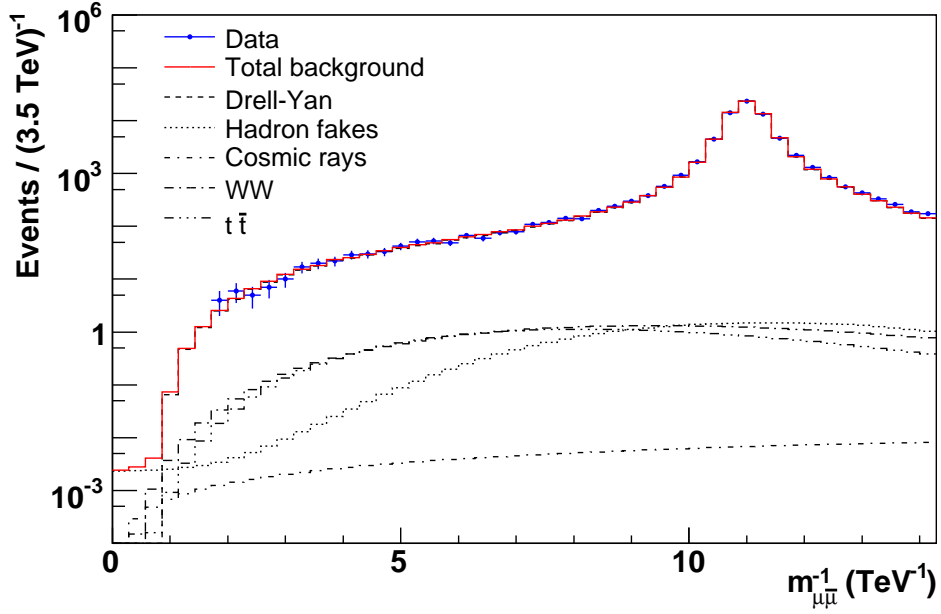,width=5.5in}}
\vspace*{8pt}
\caption{The inverse invariant dimuon mass for data and expected 
background.  The $Z$ boson peak is prominent 
at $\approx 11$ TeV$^{-1}$, and a new resonance would appear as 
a similar (narrower) peak in the sloping region on the left.
\protect\label{invmass}}
\end{figure}

\begin{figure}[!h]
\centerline{\psfig{file=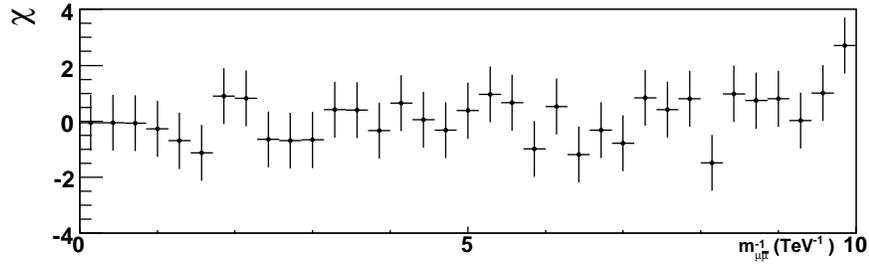,width=5.in}}
\vspace*{8pt}
\caption{The difference between data and expected background, 
divided by the statistical uncertainty.  The data are 
consistent with the background expectation; the most significant excess 
appears in the rightmost bin, corresponding to a mass of 103 GeV.
\protect\label{invmasschi}}
\end{figure}

The observed dimuon invariant mass spectrum (Fig.~\ref{invmass}) 
shows no evidence for a new neutral resonance.  This can be clearly 
seen in the error-weighted difference between observation and 
expectation (Fig.~\ref{invmasschi}).  A narrow resonance would appear 
as a significant excess in one bin, with additional excesses in neighboring 
bins.  The most significant observed excess occurs at the lowest probed 
mass in the search, 103 GeV.  An ensemble of simulated experiments gives a 
6.6\% probability of finding a more significant excess anywhere in the search 
region from a background fluctuation.

\begin{figure}[!htbp]
\centerline{\psfig{file=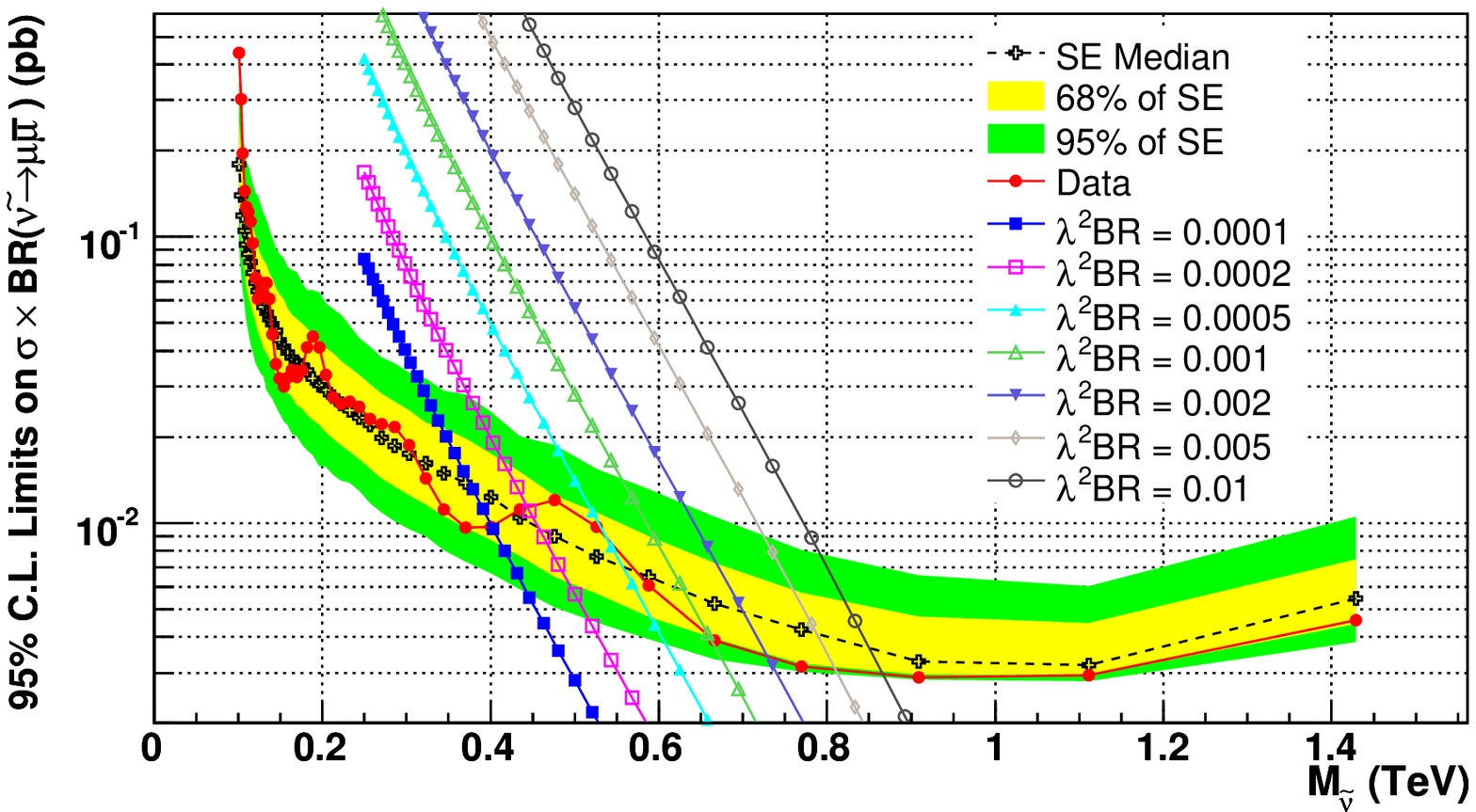,width=4.in}}
\vspace*{8pt}
\caption{The cross section limits for a new spin-0 resonance, and 
the theoretical predictions for sneutrino production for various values 
of the coupling squared ($\lambda'^2_{i11} \equiv \lambda^2$) times 
the branching ratio to dimuons.
\protect\label{snu}}
\end{figure}

\begin{figure}[!htbp]
\centerline{\psfig{file=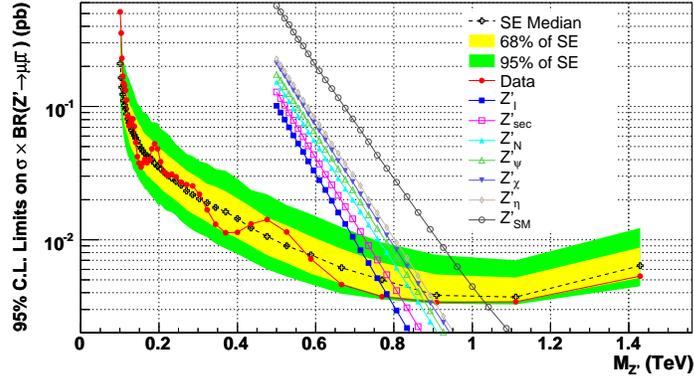,width=4.in}}
\vspace*{8pt}
\caption{The cross-section limits for a new spin-1 resonance, and the 
theoretical predictions for $Z'$ bosons with the same couplings to 
fermions as the $Z$ boson ($Z'_{SM}$), and in various E$_6$-inspired 
models. 
\protect\label{zprime}}
\end{figure}

\begin{figure}[!htbp]
\centerline{\psfig{file=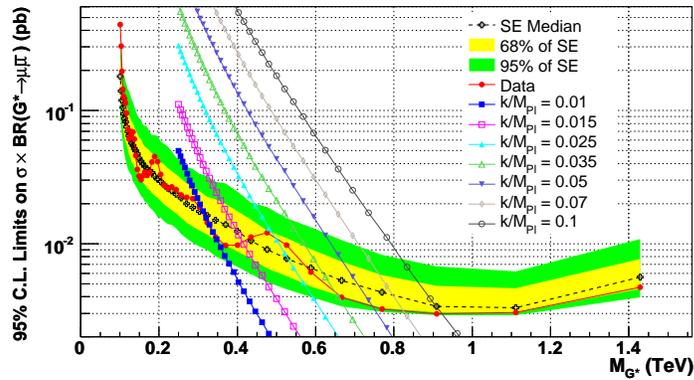,width=4.in}}
\vspace*{8pt}
\caption{The cross-section limits for a new spin-2 resonance, and the 
theoretical predictions for R-S gravitons for various values of $k/M_{Pl}$.
\protect\label{graviton}}
\end{figure}

The results are translated into cross-section limits for new resonance 
production (Figs.~\ref{snu}-\ref{graviton}), and then into mass limits 
for specific models (Table~\ref{massLimits}).  The mass limits are the 
highest of any search, except for models with weak couplings.  For such 
couplings, the probed mass range is lower and the recent CDF dielectron 
search \cite{CDFDielectron} has better sensitivity, since the CDF 
calorimeter has broader coverage in $\eta$.  The lower the resonance mass, 
the larger the average boost in the beam direction, and the larger the 
average rapidities of the leptons.  The dielectron analysis thus has 
better acceptance and sensitivity at these masses.

\begin{table}[htpb]
\caption{95\% C.L. lower mass limits, in GeV, for sneutrinos, $Z^\prime$ bosons, and 
gravitons with various model parameters.  The $Z^\prime_{SM}$ boson has the same couplings 
to fermions as the $Z$ boson. }
{\begin{tabular}{cc|cc|cc}
\hline
\hline
$\tilde{\nu}$ & $\tilde{\nu}$ & $Z^\prime$  &  $Z^\prime$    & RS graviton  &  graviton \\
$(\lambda'_{i11})^2 \cdot BR$ & mass limit & model  &  mass limit   & $k/M_{\mathrm{Planck}}$ & mass limit \\
\hline
0.0001 & 397 &  $Z^\prime_{I} $     &  789    & 0.01  & 293  \\
0.0002 & 441 &  $Z^\prime_{\mathrm sec}$ &  821   &  0.015 & 409  \\
0.0005 &  541 & $Z^\prime_{N}   $ &  861   &  0.025 & 493 \\ 
0.001 &  662 & $Z^\prime_{\psi}  $ &  878   &  0.035 & 651 \\ 
0.002 &  731 & $Z^\prime_{\chi}  $ &  892   &  0.05  & 746 \\ 
0.005 &  810 & $Z^\prime_{\eta}  $ &  904   &  0.07  & 824 \\ 
0.01 &  866 & $Z^\prime_{SM}    $  & 1030   &  0.1   & 921 \\
\hline
\hline
\end{tabular}}
\label{massLimits}
\end{table}

\subsection{CDF Dielectron Search}

The CDF dielectron search in $0.45$ fb$^{-1}$ of $p\bar{p}$ collision data is the only 
hadron-collider search for $Z^\prime$ bosons to use the dilepton angular information.  
The search parametrizes the detector response in the ($m_{ee},\cos\theta^*$) plane and 
distinguishes the SM and $Z^\prime$-boson hypotheses using the observed distribution 
in this plane.  The data are consistent with the SM so limits are set on $Z^\prime$ 
bosons in a generalized model parameter space.

\subsubsection{Two-dimensional ($m_{ee}, \cos\theta^*$) Scan}

The scattering amplitude for the $f\bar{f} \rightarrow e^-e^+$ process is \cite{angle}
\begin{eqnarray}
A_{ij} =  A(f_i \bar{f} \rightarrow e_j^- e^+) = & -Qe^2 + 
\frac{\hat{s}}{\hat{s} - m_Z^2 +im_Z \Gamma_Z}C_i^Z(f)C_j^Z(e) + \nonumber \\
&  \frac{\hat{s}}{\hat{s} - m_{Z^{\prime}}^2 + 
i m_{Z^{\prime}}\Gamma_{Z^{\prime}}} C_i^{Z^{\prime}}(f) C_j^{Z^{\prime}}(e), 
\end{eqnarray}

\noindent 
where $i$ and $j$ are the fermion helicities ($L, R$), $Q$ is the electromagnetic charge 
of fermion $f$, $C_{i,j}^{Z,Z^\prime}(f)$ are the fermion couplings to the $Z$ and $Z'$ 
bosons, $m_Z$ ($\Gamma_Z$) and $m_{Z^\prime}$ ($\Gamma_{Z^\prime}$) are the respective $Z$ 
and $Z^\prime$ boson masses (widths), and $\hat{s}$ is the squared center-of-mass energy 
of the collision.  Using this amplitude, the differential angular cross section is
\begin{equation}
\frac{d\sigma}{d\cos\theta^*} = \frac{1}{128\pi\hat{s}}
[(|A_{LL}|^2 + |A_{RR}|^2)(1+\cos\theta^*)^2 + (|A_{LR}|^2 + |A_{RL}|^2)(1-\cos\theta^*)^2].
\end{equation}

\noindent
A $Z^\prime$ boson with non-zero couplings to quarks and electrons alters the SM 
$\cos\theta^*$ distribution.

The CDF search probes the ($m_{ee},\cos\theta^*$) plane in (10 GeV, 0.25) bins using a 
look-up table from the full detector simulation to determine the acceptance in each bin.  
The $Z'$ boson signal hypothesis, including interference with the Drell-Yan process, is 
compared to SM $Z/\gamma^*$ boson production through a likelihood ratio for the two 
hypotheses in the search region of $m_{ee} > 200$ GeV.  The processes are modelled with 
{\sc pythia} \cite{pythia} and a mass-dependent NNLO correction factor \cite{nnlo}. 

In the search region CDF estimates the following SM backgrounds in $0.45$ fb$^{-1}$ of 
data:  80 Drell-Yan events; 28 events with a jet misreconstructed as an electron (dijet 
and $W$ + jet); and 7 diboson ($WW$ and $WZ$) events.  The misreconstructed-jet events 
are estimated by applying a jet-to-electron misreconstruction rate to all jets in events 
with one reconstructed electron and at least one jet.  Diboson events are estimated with 
{\sc pythia} and their theoretical cross sections \cite{ww}. 

\subsubsection{Results}

\begin{figure}[!h]
\centerline{\psfig{file=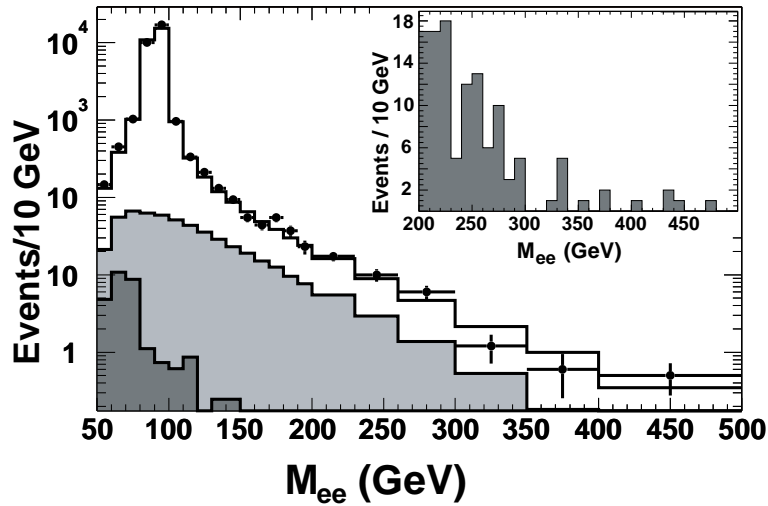,width=4.in}}
\vspace*{8pt}
\caption{The invariant dielectron mass for data and expected background.  
The inset shows the data events in the search region $m_{ee} > 200$ GeV. 
\protect\label{mass}}
\end{figure}

\begin{figure}[!h]
\centerline{\psfig{file=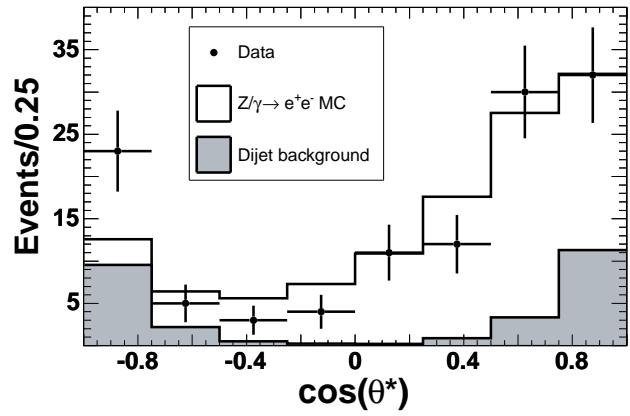,width=3.25in}}
\vspace*{8pt}
\caption{The $\cos\theta^*$ distribution in $m_{ee} > 200$ GeV search region. 
\protect\label{costheta}}
\end{figure}

The 120 observed events in the search region are consistent with the $115^{+16}_{-19}$ 
expected background events in the two-dimensional ($m_{ee},\cos\theta^*$) plane.  The 
projections along the $m_{ee}$ and $\cos\theta^*$ axes are shown in Figs.~\ref{mass} and 
\ref{costheta}.  

\begin{figure}[!h]
\centerline{\psfig{file=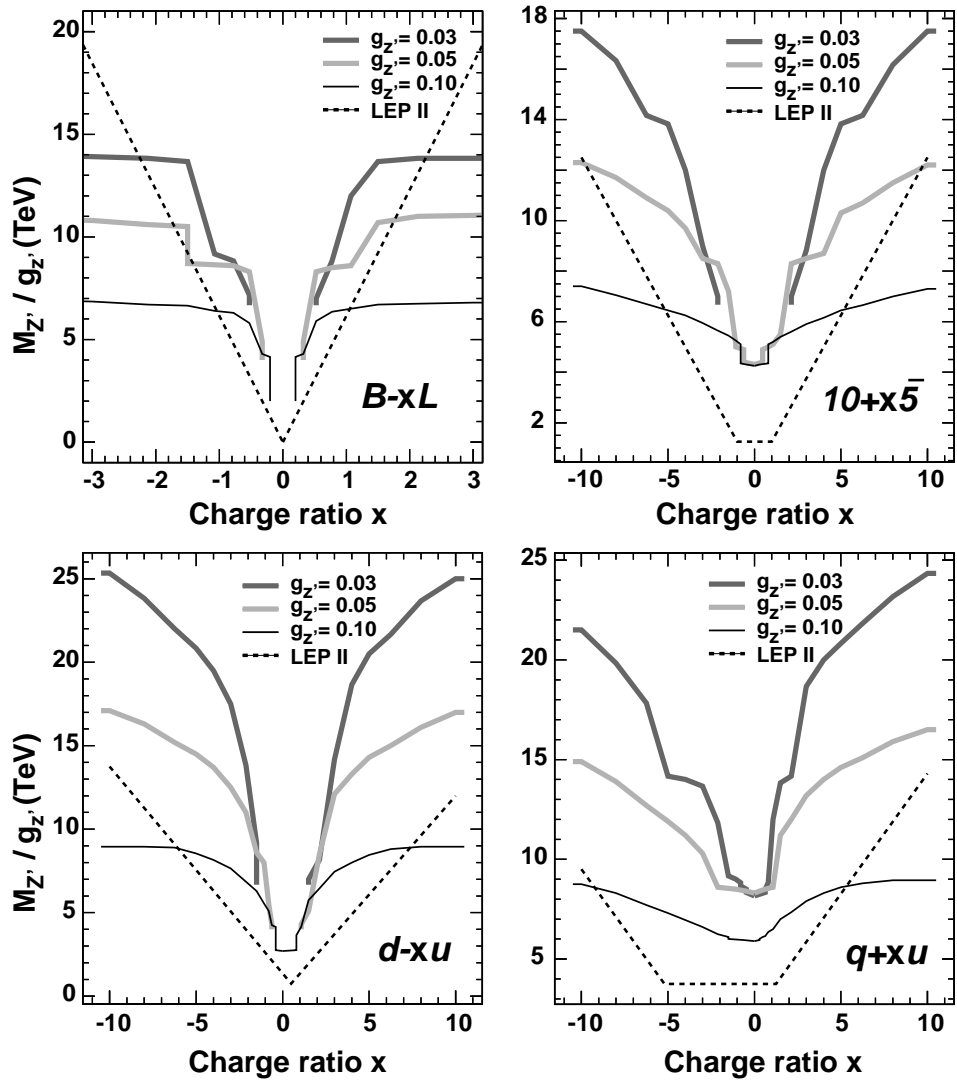,width=5.in}}
\vspace*{8pt}
\caption{Limits from the CDF dielectron search in the generalized 
($x,M_{Z^\prime} / g_{Z^\prime}$) plane for several values of $g_{Z^\prime}$.  
Values below the curves are excluded, and LEP II limits are shown for 
comparison. 
\protect\label{genZprime}}
\end{figure}

From the comparison of SM and $Z^\prime$ hypotheses, limits are set on the masses and 
couplings of $Z^\prime$ bosons.  The mass limits range from 675 GeV for $Z^\prime_{sec}$ 
to 860 GeV for $Z^\prime_{SM}$.  For the generalized models $B-xL$, $q+xu$, $d-xu$, and 
$10+x\bar{5}$, limits are set in the two-dimensional plane of 
($x, M_{Z^\prime} / g_{Z^\prime}$) for several values of $g_{Z^\prime}$ 
(Fig.~\ref{genZprime}).  The limits extend to smaller values of $|x|$ and $g_{Z^\prime}$ 
than those from LEP II.

\section{Future Searches}

The full Run II data sets are expected to be a factor of 4 larger than that used in 
the CDF dimuon analysis.  The highest current mass limits are at the edge of a steeply 
falling parton-luminosity curve and will not significantly increase.  However, there is 
ample opportunity to see hints of a weakly-coupled new resonance at a mass below the 
kinematic threshold using the full CDF and D0 data sets.  With the higher energy 
collisions soon expected from the LHC, the highest probed mass could increase by a 
factor of 3 or more \cite{lhc}.  Thus, neutral resonance searches will continue to 
provide significant discovery potential into the future.  

The technique applied to the CDF dimuon search can be extended to searches in other 
final states.  For example, at high mass the calorimetric measurement of electrons and 
photons has a fractional energy resolution that is constant in energy because the 
calorimeter sampling term becomes negligible in comparison.  Since $\delta \ln m = \delta m / m$, 
a new resonance decaying to electron pairs will have a constant peak width in the $\ln m$ 
distribution.  This is not affected by the fractional intrinsic width, which is also constant 
in mass.  For resonances decaying to quarks and gluons, the optimal distribution depends on 
the calorimeter.  At low energy the resolution is proportional to $\sqrt{E}$, so the fractional 
resolution improves with increasing energy.  In this case resonances will have constant width 
in the $\sqrt{E}$ distribution.  At sufficiently high mass, however, the intrinsic width and 
constant fractional resolution term will become dominant, in which case resonances will have 
constant width in $\ln m$.

The angular distributions of the final-state particles included in the $Z^{\prime}$-boson search 
at CDF have also been used by D0 to search for graviton production \cite{gravitonint,d0graviton}.  
Optimally, an unbinned likelihood can be performed for each spin hypothesis, using the full 
matrix-element and resolution information on an event-by-event basis.  This technique has been 
applied to Higgs-boson and top-quark searches \cite{diw,mesearch} and measurements \cite{memeasure} 
at the Tevatron.  For a high-mass resonance search, a modest gain in sensitivity is expected 
beyond a two-dimensional $(m_{ll},\cos\theta^*)$ fit.

\section*{Acknowledgments}

We would like to recognize the Tevatron accelerator division and the CDF and D0 Collaborations, 
who have significantly extended the sensitivity to new neutral resonances over the past decade.


\begin{thebibliography}{0}
\bibitem{units} We use natural units where $c \equiv \bar{h} \equiv 1$ throughout.
\bibitem{conventions} CDF uses a cylindrical coordinate system in which $+z$ points in the 
direction of the proton beam (east), $\phi$ is the azimuthal angle and $r$ is the radius 
from the center of the detector.  The rapidity $y = -\frac{1}{2}\ln[(E - p_z)/(E + p_z)]$ 
is additive under Lorentz boosts along the $z$ axis and reduces to the pseudorapidity 
$\eta = -\ln[\tan(\theta/2)]$ for massless particles, where $\theta$ is the polar 
angle with respect to the $z$ axis.
\end{thebibliography}
\end{document}